\documentclass[a4paper,12pt]{article}
\usepackage[cp1250]{inputenc}
\usepackage{epsfig}
\usepackage{amssymb}
\usepackage[english]{babel}
\usepackage{amsmath}
\usepackage{color}
 0
\begin{document}
\title{Green function diagonal for a class of heat equations}
\author{Grzegorz Kwiatkowski, Sergey Leble\\Gdansk University of Technology, \\{\small ul. Narutowicza 11/12, 80-952, Gdansk, Poland}}
\maketitle
\date{\today}

\begin{abstract}
A construction of the heat kernel  diagonal is considered as element of generalized Zeta function, that, being meromorfic function, its gradient at the origin defines determinant of a differential operator in a technique for regularizing quadratic path integral. Some classes of explicit expression  in the case of finite-gap potential coefficient of the heat equation are constructed.

\end{abstract}

\section{Introduction}
A necessity of a Green function diagonal study is directly connected with the generalized zeta-function (GZF) theory of elliptic differential operators \cite{S},
 which is successfully applied to a regularization of the operators determinants \cite{H}. Such elliptic problems, for example, appear as Laplace transform of heat kernel equations, generally,  with variable coefficients, which are conventially named as potentials. An important application of the theory is evaluation of semiclassical quantum corrections calculations to nontrivial classical solutions of important nonlinear equations and field theory \cite{RS,Kon}. The corrections are intimately linked to the fundamental solutions of related linear problems for the heat operator Laplace transform, which diagonal enters the zeta-function definition. Such regularization, for example, is realized in explicit form for kink solutions of the integrable Sine-Gordon equation \cite{kwant,ZL}
as well as  non-integrable Landau-Ginzburg ($\phi^4$) models \cite{Kon,ZL}. The kink solution (as well as multikink one) in this context corresponds  to the case of point spectrum of the elliptic operators that appear after division of variables.

This paper is devoted to investigation of a wide class of potentials which spectrum is continuous  with  eventual gaps - more precisely so-called finite-gap ones - see, e.g. the book \cite{BEB}.

Such potentials and, especially three-gap one, correspond to basic three-wave interaction, which is important in many quasiperiodic processes description, an exemplary applications one can find in \cite{BB}.

In the Sec. 2, starting from Laplace transform of the heat equation by time,  we derive a nonlinear equation for the Green function diagonal, along ideas similar ones, mentioned in \cite{C.H, BEB} in the context of other equations. We construct its solutions in cases which potentials has direct link to the polynomial functions in appropriate variables (Sec. 3). The last section is devoted to examples and appendix contains a Mathematica program, related to a class of illustrations.

\section{The equation}
We are interested in a class of problems, connected with the parabolic partial differential operator Green function (kernel of heat equation)
\begin{equation}
	\left(\frac{\partial}{\partial y}+\frac{\partial^2}{\partial x^2}-U(x)\right)g(x,x_0,y)=\delta(x-x_0)\delta(y),
\end{equation}
 where $g(x,x_0,y) \in S$ is the fundamental solution over Schwartz space $S$; $\delta(x-x_0), \quad \delta(y)$ are Dirac delta-functions.
After Laplace transform:
\begin{equation}\label{a}
	\left(p+\frac{\partial^2}{\partial x^2}-U(x)\right)\hat{g}(x,x_0,p)=\delta(x-x_0)
\end{equation}

The construction of GZF in fact rely upon the Green function diagonal. 
In \cite{ZL} a statement about
 $\hat{g}(p,x,x)=G(p,x)$,
 is used. Namely,
 $G(p,x)$ solves the equation
\begin{equation}\label{Hermit}
    2GG''_{xx} - (G'_x)^2 - 4(U(x)-p)G^2+1=0.
\end{equation}
on condition, that $U(x)$ is bounded. The equation resembles one of derived by Hermit for \cite{C.H}.

\textbf{Proof:}

Let us consider homogeneous equation
\begin{equation}\label{aa}
    \left(p + \frac{\partial^2}{\partial x^2} -U(x)\right)f\left(p,x,x_{0}\right)=0.
\end{equation}

The fundamental solution of (\ref{aa}) is built by standard procedure \cite{MW}. It has two linearly independent solutions, for example $\phi$ and $\psi$, converging respectively at $-\infty$ and $+\infty$.
One can represent $\hat{g}_{D}$ through $\phi$ and $\psi$ respectively for $x<x_0$ and $x>x_0$ with a sewing condition determined by equation (\ref{a})
\begin{equation}\label{b}
    \hat{g}_{D}(p,x,x_0)=\left\{\begin{array}{c}
                      A(x_0)\phi(p,x),\ x\leq x_0 \\
                      B(x_0)\psi(p,x),\ x\geq x_0
                    \end{array}\right. .
\end{equation}
From continuity condition of $\hat{g}_{D}$ one gets
$$A(x_0)\phi(p,x_0)=B(x_0)\psi(p,x_0).$$
What leads to:
$$A(x_0)=C(x_0)\psi(p,x_0),$$
$$B(x_0)=C(x_0)\phi(p,x_0).$$
Due to the symmetry of Green function in respect to exchanging $x$ and $x_0$, $C(x_0)$ is constant (later referred as C). To obtain condition for derivatives of $\phi$ and $\psi$ one integrates (\ref{a}) over $x$ in an $\varepsilon$ neighbourhood of $x_0$:
\begin{equation}\label{c}
    \int_{x_0-\varepsilon}^{x_0+\varepsilon} \left(p + \frac{\partial^2}{\partial x^2} -U(x)\right)\hat{g}_{D}\left(p,x,x_{0}\right)dx=1,
\end{equation}
$$\left.\frac{\partial \hat{g}_{D}}{\partial x}\left(p,x,x_{0}\right)\right|_{x=x_0-\varepsilon}^{x_0+\varepsilon}+
\int_{x_0-\varepsilon}^{x_0+\varepsilon} \left(p-U(x)\right)\hat{g}_{D}\left(p,x,x_{0}\right)dx=1,$$
$$\frac{\partial \phi}{\partial x}(p,x_0+\varepsilon)\ C\psi(p,x_0)- \frac{\partial \psi}{\partial x}(p,x_0-\varepsilon)\ C\phi(p,x_0)+
\int_{x_0-\varepsilon}^{x_0+\varepsilon} \left(p-U(x)\right)\hat{g}_{D}\left(p,x,x_{0}\right)dx=1.$$
In $\varepsilon \rightarrow 0$ limit above equation reduces to
\begin{equation}\label{d1}
    \frac{\partial \phi}{\partial x}(p,x_0)\ C\psi(p,x_0)-\frac{\partial \psi}{\partial x}(p,x_0)\ C\phi(p,x_0)=1.
\end{equation}
Since solutions of (\ref{aa}) are linear, one can assume $C=1$.
Then (\ref{d1}) reduces to
\begin{equation}\label{e}
    \frac{\partial\phi}{\partial x}(p,x_0)\ \psi(p,x_0)=\frac{\partial\psi}{\partial x}(p,x_0)\ \phi(p,x_0)+1.
\end{equation}

Actual proof will be made, by inserting (\ref{b}) to (\ref{Hermit}). For brevity function arguments will be omitted and $'$ will denote a derivative with respect to $x$
$$2\psi \phi \left(\psi '' \phi + 2\psi' \phi' + \psi \phi '' \right)-\left(\psi '\phi+\psi \phi '\right)^2-4(U(x)-p)\psi^2 \phi^2 +1=0$$
$$2\psi^2 \phi \left(\phi''-(U(x)-p)\phi\right)+2\psi \phi^2 \left(\psi''-(U(x)-p)\psi\right)+4\psi' \phi'\psi \phi-\left(\psi '\phi+\psi \phi '\right)^2 +1=0.$$
Because of (\ref{aa}) two first elements are nullified. One also uses property (\ref{e}):
$$4\psi' \phi'\psi \phi-\left(2\psi '\phi+1\right)^2+1=0,$$
$$4\psi' \phi'\psi \phi-4\psi'^2\phi^2-4\psi'\phi-1+1=0,$$
\begin{equation}\label{g}
    \psi' \phi'\psi \phi-\psi'^2\phi^2-\psi' \phi=0,
\end{equation}
\begin{equation}
    \psi'^2\phi^2+\psi'\phi-\psi'^2\phi^2-\psi' \phi=0.
\end{equation}
Thus the proof is concluded.

It is important to note, that the transition is general and doesn't rely on the nature of $U(x)$ as long as it's bound. It's usefulness is dependent on a few qualities of the potential though.

\section{The main equation solution}
\subsection{Substitutions}
 We consider a class of solutions of the equation (\ref{Hermit}), to be written in a form
\begin{equation}\label{rep}
	G(p,x)=\frac{P(p,x)}{2\sqrt{Q(p)}}.
\end{equation}
It is most useful, if there exists a variable transition $x\rightarrow z$, $U(x)\rightarrow u(z)$, in which $P$ and $Q$ are polynomials. Basic conditions for it to be possible are:
\begin{enumerate}
	\item $u$ is a polynomial in $z$,
	\item $(z'_x)^2$ is a polynomial in $z$ (note, that $z''_{xx}=\frac{1}{2}\frac{\partial}{\partial z}(z'_x)^2$),
\end{enumerate}
This does not ensure simplicity of solutions as will be shown further in the text. At this point, it is important to notice, that the second condition restricts $z(x)$ - apart from a class of elementary functions - to elliptic and hyperelliptic functions - see, e.g. \cite{BEB}.

Let us assume, that $(z'_x)^2$ and $u$ are polynomials in the variable $z$ of degree $L+1$,  and $K$ respectively, hence, $z''_{xx}$ is a polynomial of the degree $L$. We also assume, that coefficient by the highest power term of $(z'_x)^2$ is equal to $1$ (which is always attainable). After the change of variables, the equation will take the form:
\begin{equation}
	2P(P''(z'_x)^2+P'z''_{xx})-\left(P'z'_x\right)^2-4(u(z)-p)P^2+4Q=0
\end{equation}

We also assume solution in a given form:
\begin{equation}\label{sol}
\begin{array}{c}
	P(p,z)=\sum_{n=0}^{N}p^n \sum_{l=0}^{M_n}P_{n,l} z^l\\
	Q(p)=\sum_{n=0}^{2N+1} q_n p^n
\end{array}
\end{equation}
\subsection{Classification}

We will now proceed to analyse the solution by separating the equation in respect to powers of $p$ and $z$. The equation for $p^0, z^{2M_0+max(K,L-1)}$ takes following form:

If $K>L-1$
\begin{equation}
	4u_{K}P_{0,M_0}=0
\end{equation}
If $K\leq L-1\quad \land\quad M_0\geq 1$
\begin{equation}
	2P^2_{0,M_0}(M_0 (M_0-1)+\frac{L+1}{2}M_0)-P^2_{0,M_0}M^2_0-4u_{K}P^2_{0,M_0}\delta_{K,L-1}=0
\end{equation}

\begin{equation}\label{M0}
	M_0^2+(L-1)M_0-4u_{K}\delta_{K,L-1}=0
\end{equation}
Note, that $M_0=0$ leads to $K=0$ (this case will be examined later in the text). Another conclusion is, that $K\leq L-1$ is necessary for sought type of solutions. Furthermore, this leads to following, more precise conditions:
\begin{equation}
\begin{array}{c}
	L\geq 1 \qquad (due\ to\ K\geq 0) \\
	K=L-1 \qquad (for\ M_0\ to\ have\ positive\ value)
\end{array}
\end{equation}
Equation for $p^{2N+1}$
\begin{equation}
	4\left(\sum_{l=0}^{M_N}P_{N,l} z^l\right)^2+4q_{2N+1}=0
\end{equation}
leads to following conclusions:
\begin{equation}
	M_N=0\quad \land\quad P^2_{N,0}=-q_{2N+1}.
\end{equation}
Let us now look at subsequent equations for descending powers of $p$. For $p^{2N}$ we have
\begin{equation}
	-4u(z)P^2_{N,0}+8P_{N,0}\sum_{l=0}^{M_{N-1}}P_{N-1, l}z^l+4q_{2N}=0,
\end{equation}
which leads to
\begin{equation}\label{2N}
	\sum_{l=0}^{M_{N-1}}P_{N-1, l}z^l=\frac{1}{2}\left(P_{N,0}u(z)-\frac{q_{2N}}{P_{N,0}}\right),
\end{equation}
\begin{equation}
	M_{N-1}=K.
\end{equation}
For $p^{2N-1}$ we get (for the highest power of $z$):
on condition $M_{N-2}>M_{N-1}+K$ ($z\neq x^2$)
\begin{equation}
4P_{N,0}P_{N-2,M_{N-2}}=0,
\end{equation}
on condition $M_{N-2}\leq M_{N-1}+K$
\begin{equation}
	\begin{array}{c}
	2P_{N,0}P_{N-1,M_{N-1}}(M_{N-1}(M_{N-1}-1)+\frac{K+2}{2}M_{N-1}) \\ -2P_{N,0}P_{N-1,M_{N-1}}-8u_{K}P_{N,0}P_{N-1,M_{N-1}} \\ +8P_{N,0}P_{N-2,M_{N-2}}\delta_{M_{N-2},L-1}+4q_{2N-1}=0.
	\end{array}
\end{equation}
It's obvious, that $M_{N-2}\leq M_{N-1}+K$ is a necessary condition for (\ref{sol}). Now we can consider a general rule for all remaining equations. Thesis: $\forall_{0\leq k<N-1} M_{k}\leq (N-k)K$. Proof by induction:
If $M_{k}> (N-k)K$, then equation for $p^{N+k+1}$ and highest power of $z$ takes form:
\begin{equation}
	4P_{N,0}P_{k,M_k}=0
\end{equation}
If $M_{k}\leq (N-k)K$ and $\forall_{k<l<N} M_l=(N-l)K$ (possibility giving the highest possible value of $M_k$), then equation for $p^{N+k+1}$ and highest power of $z$ takes form:
\begin{equation}
	\begin{array}{c} 2P_{N,0}P_{k+1}(M_{k+1}(M_{k+1}-1)+\frac{K+2}{2}M_{k+1})\\ +2\sum_{n=k+2}^{N-1}P_{n,M_n}P_{N-n+k+1,M_{N-n+k+1}}(M_{N-n+k+1}(M_{N-n+k+1}-1) \\ +\frac{K+2}{2}M_{N-n+k+1})  -2\sum_{n=k+2}^{N-1}P_{n,M_n}P_{N-n+k+1}M_{N-n+k+1}M_{n} \\ -4u_{K}\sum_{n=k+1}^{N}P_{n,M_n}P_{N-n+k+1} \\ +4\sum_{n=k+1}^{N-1}P_{n,M_n}P_{N-n+k,M_{N-n+k}} \\ +4\delta_{M_{k},M_{k+1}+L-1}P_{N,0}P_{k,M_k}=0
	\end{array}
\end{equation}
Thus the solution exists only if the thesis holds. This leads directly to a minimal condition on $N$:
\begin{equation}\label{Nmin}
	N\geq\frac{M_0}{K} \qquad \forall M_0\geq K
\end{equation}
In summary: existence of solutions of form (\ref{sol}) depends on the power of the potential ($K$), power of $(z'_x)^2$ ($L\leq 1$ can give abnormal results), amplitude of the highest power term of the potential and there exists a definite formula for the minimal value of $N$.
\subsection{Solving algorithm}\label{Al}
\begin{enumerate}
\item If there exists a solution in the form (\ref{sol}) for a given $N$ (which fulfils requirement (\ref{Al})), it can be obtained in a straightforward manner. Since the actual value of $N$ is unknown, one starts with the minimal possible value $\frac{M_0}{K}$. After separating the equation in respect to powers of $p$ one analyzes the resulting equations starting from the highest power of $p$. All those equations can be written in a manner similar to (\ref{2N}) (here for the $p^{N+n+1} with possible $n$ from $N-1$ to $0$)$:
\begin{equation}
	P_{N,0}\sum_{l=0}^{(N-n)*K}P_{n,l}z^l=F(P_{N,0},P_{N-1,K},P_{N-1,K-1},\dots,P{n+1,0},z)-\frac{q_{N+n+1}}{2}
\end{equation}
where $F$ contains all elements of equation not written explicitly. It is easy to see, we can obtain all coefficients, except for $P_{n,0}$, as solutions of linear equations, since all elements on the RHS, except for $q_{N+n+1}$ are known. As for the equation for $z^0$, it is more convenient, to express $q_{N+n+1}$ in terms of $P_{n,0}$. Solving all equations down to $p^{N+1}$ gives us all $P_{n,l}$ as well as some of $q_n$ in terms of $\{P_{n,0}\}_{i\in\{0,...,N\}}$. It is important to note, that all calculations done up to that point stay relevant, even if value of $N$ will have to be increased. 
\item In the next step, we use the equation for $p^N$ to calculate the possible values of $P_{n,0}$. Again, if we start from the highest powers of $z$, we can obtain those coefficients as solutions of linear equations, since any element containing $P_{n,0}$ is proportional to at most $z^{K(N-n+1)}$ and any element containing $P^2_{n,0}$ is proportional to at most $z^{K(N-2n+1)}$ (negative exponent means, that such coefficients are not present in equation for $p^N$). 
\item Subsequently we check the solution. If it doesn't hold, we increase the value of n, add relevant components to $P$ and $Q$ polynomials, calculate values of all new coefficients and go back to step 2.
\end{enumerate}
Since $q_n$ aren't necessary for calculation of any $P_{n,l}$ and checking the solution, we can slightly simplify the algorithm by calculating $q_n$ after all other coefficients.

\noindent Described algorithm was implemented in Mathematica 7 and used to obtain solutions presented in section \ref{sol}.
\subsection{On uniqueness of solutions}

Using the above algorithm we obtain all coefficients as solutions of linear equations in respect to sought coefficients. This leads to a simple conclusion, that for a given $N$ solutions of form (\ref{sol}) are unique except for constant $u$, in which case the solution has no dependence on $z$ and because of this, there is no relation between any of $P_{n,0}$ (\ref{const}). As yet, there is no method of finding all allowed $N$ for a given potential. Therefore uniqueness of solutions is uncertain.

\section{Exemplary solutions}\label{sol}
\subsection{Constant potential}\label{const}
Let's consider a constant potential
\begin{equation}\label{con}
	U(x)=u
\end{equation}
It is obvious, that no change of variables is necessary, thus we can use $z=x$ ($L=-1$). Equation for $p^0 z^{2M_0}$ immediately gives
\begin{equation}
	-4uP^2_{0,M_0} +4q_0\delta_{M_0,0}=0
\end{equation}
Since $u$ can have an arbitrary value, if we shift the $p$ variable, this equation only holds for $M_0=0$. This means, that the simplest solution would be
\begin{equation}
	G(p,x)=\frac{1}{2\sqrt{u-p}}
\end{equation}
This is not the only one, as will be shown. Let us consider a solution for potential (\ref{con}) and an arbitrary $N$. Equation for $p^{2N+1}$ gives as usual
\begin{equation}
	M_N=0
\end{equation}
\begin{equation}
	q_{2N+1}=-P^2_{N,0}
\end{equation}
Equation for $p^{2N}$ gives
\begin{equation}
	M_{N-1}=0
\end{equation}
\begin{equation}
	P_{N-1,0}=\frac{1}{2}\left(uP_{N,0}-\frac{q_{2N}}{P_{N,0}}\right)
\end{equation}
Subsequent equations will look alike, with $M_i=0$ for all $i$. It is easy to see, that one obtains a total of $3N+3$ parameters with only $2N+2$ equations and one can obtain a solutions for any value of $N$. In a sense, it's a consequence of the condition (\ref{Nmin}), since $\frac{0}{0}$ is an indeterminate symbol.
\subsection{Triple-gap cnoidal potential}
Let's take a solution one order higher then that for $\phi^4$ cnoidal solution:
\begin{equation}
	U(x)=-12m^2k^2\ cn^2(mx;k)
\end{equation}
\begin{equation}
	z=cn^2(mx;k)
\end{equation}
\begin{equation}
	\left(z'_x\right)^2=4m^2z(1-z)(1-k^2+k^2z)
\end{equation}
\begin{equation}
	z''_{xx}=2m^2(-3k^2z^2+(4k^2-2)z+1-k^2)
\end{equation}
\begin{equation}
	M_0=3
\end{equation}
\begin{equation}
	N=3
\end{equation}
Algorithm explained in section (\ref{Al}) gives (assuming $P_{3,0}=1$ for simplicity):
\small
\begin{eqnarray}
	P_{2} & = & -2m^2(7 + k^2 (-14 + 3 z)) \\
	P_{1} & = & m^4(49 + k^2 (-256 + 78 z) + k^4 (256 + 3 z (-52 + 15 z))) \\
	P_{0} & = & -3m^6(12 + 8 k^2 (-19 + 9 z) + 3 k^4 (128 + z (-121 + 45 z)) +\\ & &
   k^6 (-256 + 3 z (121 + 5 z (-18 + 5 z)))) \\
	Q(p) & = & -((-4 + 8 k^2) m^2 + p) ((9 - 96 k^2 + 96 k^4) m^4 +
   10 (-1 + 2 k^2) m^2 p + p^2)\\ & & ((9 - 42 k^2 + 33 k^4) m^4 +
   2 (-5 + 7 k^2) m^2 p + p^2)\\ & & (3 k^2 (-8 + 11 k^2) m^4 +
   2 (-2 + 7 k^2) m^2 p + p^2)
\end{eqnarray}
\normalsize
\section{Conclusion}
Described equation allows calculation of heat equation's Green function diagonal in a straightforward manner. Developed algorithm should be especially useful for finding solutions for finite-gap potentials, which naturally emerge in periodic and quasi-periodic structures.


\begin{thebibliography}{ }
\bibitem{S} R.T. Seeley, Singular integrals on compact manifolds, Amer. J. Math. 81 (1959), 658-690.  Regularization of singular integral operators on compact manifolds, Amer. J. Math. 83 (1961), 265-275. MR 123167.
\bibitem{H} Hawking, S. W. - Zeta Function Regularization of Path Integrals in Curved Spacetime
\bibitem{RS} V.N. Romanov and A.S. Schvarts, Anomalies and elliptic operators, ~ Teor. Mat. Fiz., 41, ' 190 (1979).
\bibitem{Kon} R.V. Konoplich \textit{Quantum corrections calculations to nontrivial classical solutions via zeta-function} Teor. Mat. Fiz., Vol. 73, No. 3, pp. 379-392, December, 1987. \textit{The zeta-function method in field theory at finite temperature} Teor. Mat. Fiz., Vol. 78, No. 3, pp. 444-457, 1989.
\bibitem{kwant} A. Alonso Izquierdo, W. Garc\'{i}a Fuertes, M.A. Gonz\'{a}lez Le\'{o}n, J. Mateos Guilarte \textit{Generalized zeta functions and one-loop corrections to quantum kink masses} Nuclear Physics B 635 [PM] 525-557, 2002
\bibitem{ZL} Anatolij Zaitsev, Sergey Leble \textit{Quantum corrections to static solutions of phi-in-quadro and Sin-Gordon models via generalized zeta-function} arXiv:0804.1255v1  [quant-ph]  8 Apr 2008; in: Nonlinear Dynamics Editor(s): M. Daniel, S. Rajasekar, Narosa publishing house, ISBN: 978-81-7319-941-7 Publication Year:   2009
\bibitem{BEB} Belokolos, E.D., Enolski, V.Z., Bobenko, A.I.,Its, A.R., and  Matveev,
V.B. (1994) \textit{Algebro-Geometric Approach to Integrable Differential Equations}, Springer-Verlag, Berlin.
\bibitem{C.H} C. Hermite: Ouvres. Sur l'\'{E}quacion de Lam\'{e}, p. 118.  Gauthier-Villard Paris, 1912.
\bibitem{BB} Babich M., Bordag L. Quasi-periodic  vortex structures in two-dimensional flows in an inviscid incompressible fluid. Russion Journal of mathematical physics, v 12, 2005, p121-156.
\bibitem{Tyu} A. N. Tyurin, Quantization, Classical and Quantum Field Theory and Theta Functions (Am. Math. Soc., Providence, RI, 2003), CRM Monogr. Ser. 21.

\end{thebibliography}
\end{document}